\documentstyle[11pt,epsf]{article}

\def\bgg{\begin{eqnarray}\begin{array}{rcl}\displaystyle}
\def\egg{\end{array} &  & \end{eqnarray}}
\def\barr{\begin{equation} \begin{array}{*{4}{c}} }
\def\earr{\end{array} \end{equation}}

\def\schi{\raisebox{0.2em}{$\chi$}}

\newcommand{\mas}[2]{$\ ^{\scriptstyle #1} #2$}

\def\sech{\mbox{sech}}

\begin{document}

\title{Cubic Defects: Comparing the
Eight-State-System with its Two-Level-Approximation}

\author{Peter Nalbach, Orestis Terzidis\\
Institut f\"ur theoretische Physik\\
Philosophenweg 19, 69120 Heidelberg, Germany}
\date{\today}
\maketitle

\begin{abstract}
Substitutional defects in a cubic symmetry (such as a
lithium defect in a KCl host crystal) can be modeled
appropriately by an eight-state-system. Usually this
tunneling degree of freedom is approximated by a
two-level-system. We investigate the observable
differences between the two models in three
contexts. First we show that the two models predict
different relations between the temperature dependence
of specific heat and static susceptibility. Second we
demonstrate that in the presence of external forces
(pressure and electric field) the eight-state-system
shows features that cannot be understood within the
framework of the two-level-approximation. In this
context we propose an experiment for measuring the
parameter for tunneling along the face diagonal.
Finally we discuss the differences between the models
appearing for strongly coupled pairs. Geometric
selection rules and particular forms of asymmetry
lead to clear differences between the two models.
\end{abstract}

\section{Introduction}
 
Quantum tunneling of substitutional defect ions in
alkali halide crystals leads to particular low
temperature properties \cite{Nara}. Due to their misfit
in size or shape such defect ions are confined to a 
potential energy landscape with a few degenerate
potential wells. At low temperatures thermally
activated crossing over the barriers is inhibited  and
the defect ion passes through the barrier by quantum
tunneling; typically at a few Kelvin hopping becomes
relevant. The potential energy landscape in which the
defect ion moves is given by the host crystal and
therefore reflects its symmetry which for most
alkali halide crystals is cubic (e.g. the
fcc-structure of potassium chloride). There are only
three multi-well potentials which are consistent with
this  symmetry: twelve wells at the edges of a cube,
six wells in the middle of the  surfaces and eight
wells at the corners of a cube. In all cases the edges
of the cube lie along the crystal axes and the 
multi-well structure leads to off-center-sites for the
defect ion. The off-center position has two immediate
consequences: it separates the  centers of charge and
leads to a local distortion of the crystal. Hence both
an  electric dipole moment and an elastic quadrupole
moment is connected to the  defect which can thus
interact with lattice vibrations, external fields or 
neighboring defects. Consequently only at low defect
concentrations one can  describe the situation by
isolated tunneling systems. With rising concentration 
pairs, triples etc. of defects are involved until
finally one faces a  complicated many body system
\cite{Aloisbuch,Orestispaper}.
\par
A standard example of tunneling defects is potassium
chloride doped with a  small amount of lithium ions
(KCl:Li). This system is well described by  isolated
defects for concentrations up to say 20 ppm. The
minima of the system lie on the corners of a cube ($d
\approx 1.4$ \AA);  in the low temperature regime the
relevant degree of freedom is thus an 
eight-state-system (ESS). There are three different
matrix elements for  tunneling: (i) along the edges of
the cube $k$, (ii) along a face diagonal $f$ and (iii)
along a space diagonal $r$; edge tunneling dominates
the defect spectrum \cite{Gomez} for simple geometric
reasons: the edge is the shortest distance  between the
potential minima. Neglecting face and space diagonal
tunneling the  problem factorizes into three
two-level-systems (TLS); this much simpler model is
often used for the description of the defect.

In this paper we want to study in how far the TLS 
is a good approximation for 
the ESS (and hence in how far tunneling along the
face and space diagonal on one hand and the particular
geometry of the defect on the other is  negligible). We
find that in most contexts the TLS approximation is
indeed acceptable; still there are quite some
experimentally observable features which cannot be
explained by the TLS approximation. 

The plan of the paper is as follows: In section
2 a tensorial Hamiltonian for the eight-state-system
is introduced. Physical properties such as the
specific heat and the static dielectric 
susceptibility are derived and compared with the
results of the two-level-approximation. In section 3
we discuss the coupling of the eight-state-system to
static external electric or strain fields. We propose
an experiment in order to measure directly the
parameter for tunneling along a face diagonal. (This
parameter will dominate all corrections to the TLS
results.) In section 4 strongly coupled pairs of
eight-state-systems are considered. We discuss the
situation by means of group theory and investigate in
particular the relevance of different asymmetry terms
for echo experiments. Finally section 5 gives a
conclusion.

\section{Specific Heat and Static Susceptibility}

\subsection{Theory}

Before going into a discussion of the ESS we would
like to summarize some of  the well-known features of
a TLS. This system describes a particle in a
double-well-potential depending on one  single
coordinate (instead of the three space dimensions of
the ESS). The tunneling Hamiltonian of the TLS reads
in the 'local' basis (i.e. the basis  where $\sigma_z$
represents the position 'left' or 'right' of the
particle) %
\begin{equation} H = k_0 \sigma_x .\end{equation}
The symbols $\sigma_x, \sigma_z$ denote the Pauli
matrices. The only parameter  entering the system is
the tunneling parameter $k_0$ between the left and
the  right state. It is given by the WKB-formula 
\begin{equation}\label{wkb} k_0 = - E_0
\exp\left\{-\frac{d}{2 \hbar} \sqrt{2 m  V_0}\right\}
< 0 .
\end{equation}
Here the energy $E_0$ is the oscillator frequency of
a well, $d$ is the  distance between the two wells,
$m$ is the mass of the tunneling particle and 
$V_0$ is the barrier height. Since $\sigma_z$
represents the position operator, all external fields
couple  to the product $F\sigma_z$ where $F$ stands for
the amplitude of an external field (such as static
electric or strain fields and electromagnetic or
acoustic waves).

Two characteristic features of the TLS have been
studied in numerous  experiments: the temperature
dependence of the specific heat  (Schottky-anomaly)
\begin{equation}\label{cvzns} c_v(T) =
\frac{k_0^2}{k_B T^2} \quad  
\sech^2(\beta k_0 ) \end{equation}
and the temperature dependence of the static
susceptibility
\begin{equation}\label{schizns} \schi_{stat}( T ) =
\frac{  p^2}{\hbar\epsilon_0} \frac{1}{k_0} 
\tanh(\beta k_0 ). \end{equation}
Let us now ask: What observable differences appear
if we consider the ESS? In order to find an answer we
derive the corresponding formulas of this system and
compare them to those of the TLS. For the ESS three
coordinates determine the wave function in the local
basis $\{\,|xyz\rangle\,\}$ with
$x,y,z=\pm 1$; the origin of the system  is located in
the center of the cube and the axes point along its
edges. The position operator now becomes a vector
\begin{equation} \hat{\vec{r}} = \frac{d}{2} \left(
  \begin{array}{*{1}{r@{\: \;}}r@{\:}r}           
         r_{x} \\ r_{y} \\  r_{z}    
  \end{array}   \right)                 
          = \frac{d}{2} \left( 
  \begin{array}{*{1}{r@{\: \;}}r@{\:}r}       
                      \sigma_z \otimes 1\otimes 1    \\  
                    1\otimes \sigma_z \otimes 1    \\  
                  1\otimes 1\otimes \sigma_z    
  \end{array}  \right) ,
\end{equation}
where the symbol $\otimes$ denotes a tensor product
and each factor stands for  one coordinate. The
tunneling Hamiltonian of the ESS can be decomposed
into  tensor products; edge tunneling along the
$x$--direction for example is  described by
$k(\sigma_x \otimes 1\otimes 1)$. Tunneling along the
face and space diagonal involves more than one
coordinate;  the corresponding Hamiltonian is given by
\begin{eqnarray}
\label{hnull} 
\hat{H}_0=\; \;
&k& (1\otimes1\otimes\sigma_x+
1\otimes\sigma_x\otimes1+
\sigma_x\otimes1\otimes1)\nonumber\\
+&f& (1\otimes\sigma_x\otimes\sigma_x+
\sigma_x\otimes1\otimes\sigma_x+
\sigma_x\otimes\sigma_x\otimes1)\\
+&r& (\sigma_x\otimes\sigma_x\otimes\sigma_x).
\nonumber 
\end{eqnarray}
Here $k,f,r$ are the amplitudes for tunneling along an
edge, a face diagonal and a space diagonal. Let us
next write down the elastic and electric  momenta
connected with the defect. The separation of charge
which arises from the  off-center-position causes an
electric dipole moment 
\begin{equation}\label{dip} 
\vec{p} = q \vec{r},\qquad p=\frac{\sqrt{3}}{2}qd. 
\end{equation} 
The corresponding interaction energy with an external
electric field $\vec{F}$  reads
\begin{equation} 
W_F = -\sum_i F_ip_i . 
\end{equation}
Moreover the defect distorts the host crystal locally
producing thus an elastic  moment. W\"urger
\cite{Aloisbuch} derived as the leading contribution a
quadrupole moment  \begin{equation}\label{qua} Q_{ij} =
\frac{4}{d^2} r_i r_j (1 - \delta_{ij}). 
\end{equation} Remember here that $r_i$ is a component
of the position operator. The corresponding interaction
energy with a strain field is given by
\begin{equation} 
W_{\epsilon} = -g\,\sum_{i,j}Q_{ij} \varepsilon_{ij},
\end{equation} 
where $g$ is a coupling constant and
$\varepsilon_{ij}$ is the tensor of distortion
produced by the strain field. In the static case this
tensor can be derived from the exerted pressure using
standard elastomechanical relations  \cite{Landauela};
in the case of acoustic waves it reads
$\varepsilon_{ij}=1/2(\partial_j u_i + \partial_i u_j )$, 
where $\vec{u}$ denotes the amplitude of the phonon. 

Just as for the TLS the tunneling parameters can be
found with the  WKB-formula (\ref{wkb}). Assuming
that the heights of all potential barriers  are of the
same order of magnitude the three tunneling parameters
differ only  because of a simple geometric reason:
the distance separating the minima is  different (the
edge of a cube is smaller than a face  diagonal etc.).
One concludes: $|k| > |f| > |r|$. Hence in a first
approach one  can neglect $f$ and $r$. Then the
Hamiltonian (\ref{hnull}) factorizes and  three
independent TLS remain, one for each spatial
direction. This is nothing else than the
two-level-approximation keeping in mind a factor
of three:  one ESS (with $f=r=0$) is equivalent to
three TLS. One can then conclude: as soon as $f$ and
$r$ are not negligible differences between the ESS and
the TLS will arise.

Let us go further in our analysis of the ESS. The
simple structure of the  problem allows the exact
calculation of its spectrum and eigenvectors. They
are  given by the following scheme:
\begin{eqnarray}\label{mar500}   
\begin{array}{*{7}{c@{\:\;}}c@{\:}c} 
& E_3 & = & -3k+3f-r & : & | - - - \rangle  & \\
&E_2&=& -k-f+r & : & 
| + - - \rangle  ,  
| - + - \rangle  , 
| - - + \rangle  &  \\
&E_1&=&k-f-r & : &
| + + - \rangle , 
| + - + \rangle , 
| - + + \rangle  &  \\
&E_0&=&3k+3f+r & : & 
| + + + \rangle  & 
\end{array}
\end{eqnarray}
with
\begin{eqnarray}
|\pi_x\pi_y\pi_z\rangle&=&
\frac{1}{\sqrt{8}}\sum_{x,y,z=\pm 1} 
f_x^{\pi_x} f_y^{\pi_y} f_z^{\pi_z} | x y z \rangle \\ 
f_j^{\pi_j}&=& \exp \left(
\frac{i \pi}{4} (1 - \pi_j)(1 - j) \right) . 
\end{eqnarray}
Here $\pi_i = \pm$ denotes the parity of the wave
function in the $i$--direction. The spectrum consists
of four levels with a threefold  degeneracy of the
second and the third level. The transitions shown in
figures 1a and 1b are derived from the electric and acoustic
interactions given above.
As an explicit example we consider an electric field
oscillating in the y--direction.
Figure 1a gives a complete scheme of the possible transitions.

\begin{figure}[h]
\leavevmode
\centering
\epsfxsize=7cm
\epsffile{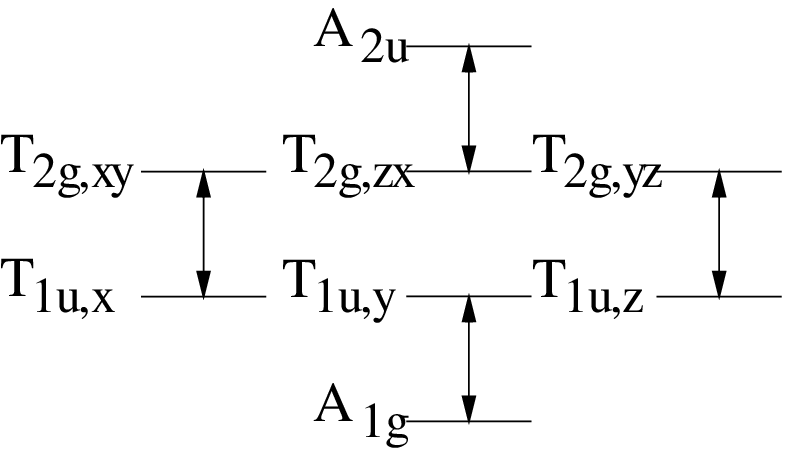}

Figure 1a: Selection rules of the ESS for an electric field
oscillating in [100]--direction together with the Schoenflies
notation.
\end{figure}

Every geometric constellation has its specific selection rules;
we refrain from giving a complete list of
all these possibilities and instead propose in figure 1b 
a schematic diagram where all possible transitions are shown. 
Note that for the electric transition one of the 
parities $\pi_i$ changes its sign whereas for the acoustic
transitions two parities change their signs.
Hence in the accoustic case the product $\pi_1\pi_2\pi_3$ 
is conserved.

\begin{figure}[h]
\leavevmode
\centering
\epsfxsize=7cm
\epsffile{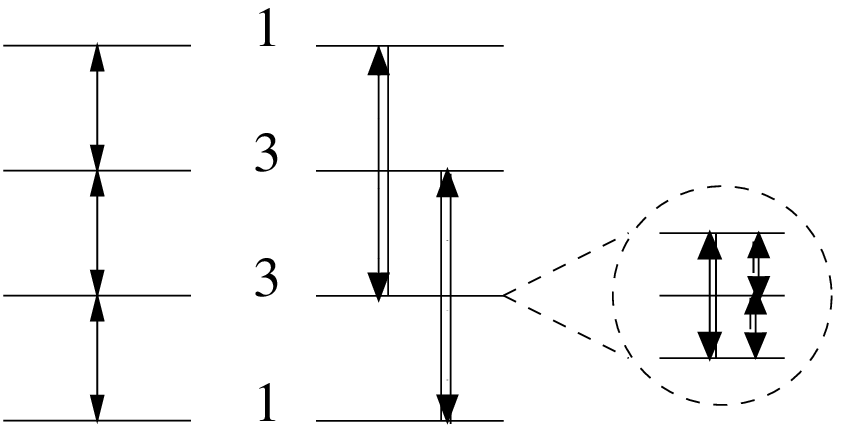}

Figure 1b: Possible electric (single arrows) and acoustic (double
arrows) selection rules for the ESS
\end{figure} 

There are three electric transitions with almost the
same frequency $\sim 2k$. In the acoustic case there
are two transitions with frequency $\sim 4k$ and 
two 'inside' the degenerate tripletts. (These
transitions are visible as soon as some perturbation
destroys the degeneracy, cf. section 3). The fact
that for both couplings only one transition frequency
appears is  crucial for the validity of the
two-level-approximation: by construction the  TLS
would not be able to reflect more than one single
energy.  Indeed the approximation cannot account for
the fact that the frequency induced  by electric
fields differs from that induced by acoustic fields. 

Let us now turn to the temperature dependence of the
specific heat and the  static susceptibility. The
partition function is given by
\[  Z = \mbox{Tr}\left(e^{-\beta H}\right) = 
\sum_l\eta_l e^{-\beta{\cal{E}}_l}\quad 
,\quad \beta=\frac{1}{k_BT}  \]
where $\eta_l$ denotes the degree of degeneracy of
the eigenvalue ${\cal{E}}_l$  and $k_B$ is the
Boltzmann constant. The free energy of the system is
then given by 
$F = - k_B T \ln Z$ and the specific heat reads
\begin{equation} 
c_v(T) = \frac{k_B\beta^2}{Z^2} 
\sum_{i<j} \eta_i\eta_j  ({\cal E}_i - {\cal E}_j)^2
e^{-\beta({\cal E}_i+{\cal E}_j)}. 
\end{equation}
Inserting the energies given in (\ref{mar500}) one
finds for the specific heat of the ESS:
\begin{eqnarray}\label{cvazs} 
c_v(T) &=& \frac{k_B\beta^2}{Z^2}
\Big\{(6k+2r)^2e^{-6\beta f}
+3(4k-4f)^2e^{-\beta(-2k+2f-2r)}\nonumber\\
&&+9(2k-2r)^2e^{2\beta f}+
3(2k-4f+2r)^2e^{-\beta(-4k+2f)} \\
&&+3(2k+4f+2r)^2e^{-\beta(4k+2f)} 
+3(4k+4f)^2e^{-\beta(2k+2f+2r)}\Big\}.\nonumber
\end{eqnarray}
This expression reduces to the corresponding formula of
the TLS when setting $f=r=0$. 
\par
Let us now write down the susceptibility $\chi$. The
response of the ESS to an external field is given by
the commutator formula \cite{Kubo} 
\[ 
\schi_{ij} = \frac{i}{\hbar} \langle [p_i(t),p_j(t_0)]
\rangle \Theta(t-t_0)  
\]
with the time-dependent dipole operator
$p(t)=\exp(i/\hbar \hat{H}t)\hat{p}\exp(-i/\hbar
\hat{H}t)$ and the Gibbs mean-value 
$\langle\cdot\rangle = \mbox{Tr}(\cdot \exp(-\beta\hat{H}))/Z$.
In order to find the static susceptibility one has to
look at the real part of  the Laplace-transform for
zero frequency. This leads to 
\begin{eqnarray}\label{schiazs} 
\chi_{x,x}^{stat}(T)=
{\displaystyle\frac{2}{Z}\,\frac{p^2}{\hbar\epsilon_0}}
\;\bigg\{ &&  
\Big[ e^{\beta(3k+3f+r)}-e^{-\beta(k-f-r)} \Big] \,
{\displaystyle\frac{1}{(2k+4f+2r)}}  \nonumber\\
&+& 2\,
\Big[e^{\beta(k+f-r)}-e^{-\beta(-k-f+r)}\Big] \;
{\displaystyle\frac{1}{(2k-2r)}}              \\
&+&
\Big[e^{-\beta(-k-f+r)}-e^{\beta(-3k+3f-r)}\Big] 
{\displaystyle\frac{1}{(2k-4f+2r)}}\bigg\}.\nonumber
\end{eqnarray}
All other elements of the susceptibility tensor
are then known: the symmetry of the problem implies
$\schi_{x,x} =
\schi_{y,y} = \schi_{z,z}$  and $\schi_{x,y} =
\schi_{x,z} = \schi_{y,z} = 0$. Again this result
reduces to  the corresponding formula for the TLS in
the case of vanishing face and space  diagonal
parameters. 

At first sight almost no contrast between the two
models emerges from the above formulas. Taking
reasonable values for $f$ and $r$ (say $r/f=f/k\approx
10-20\%$), the plots look practically the same. The
question arises whether there is at all a simple
observable feature which  distinguishes between the
TLS and the ESS. Indeed such a criterion exists. It
is based on the observation that for the TLS the
relation $k_0^2\partial_T\chi=c_V$ holds, whereas
this is not true for the ESS. One simple consequence
is that the turning point of the susceptibility and the
maximum of the specific heat {\it coincide} for a
TLS, whereas they occur at {\it different}
temperatures for the ESS. To be more explicit:
Defining   
\begin{equation}
\begin{array}{lcl} 
T_{max}:&\frac{\partial}{\partial T}
c_V(T=T_{max})&=0 \\ 
T_{turn}:&\frac{\partial^2}{\partial T^2}
c_V(T=T_{turn}) & = 0
\end{array}
\end{equation}
one finds for the TLS
\[ T_{max} = T_{turn}, \]
whereas for the ESS
\begin{equation} T_{max} < T_{turn} \end{equation}
holds.
It is not possible to give an explicit expression 
of $\Delta T = T_{turn}-T_{max}$ as a function
of $f$ and $k$ since transcendental equations are 
involved.
Instead we plot in figure 2 $\Delta T$ as a function of $f$ 
with parameter $k$.
It turns out that $\Delta T$ depends only very 
weakly on $k$ for $0.5K<|k|/k_B<1.5K$ and that a linear relation
$\Delta T \approx 0.85\; |f|/k_B$ holds approximately.

{\leavevmode
\centering
\epsfxsize=8cm
\epsffile{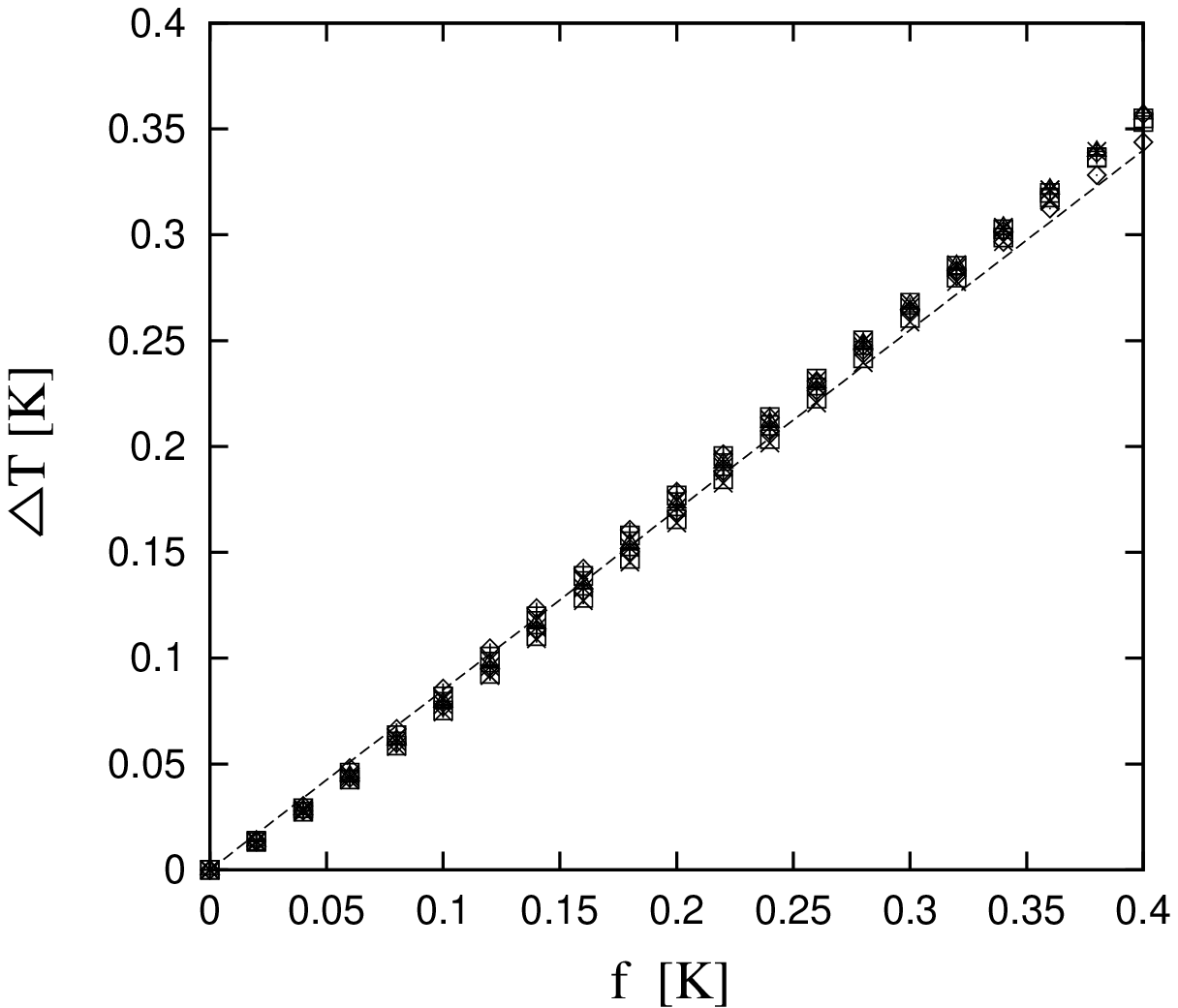}

Figure 2: $\Delta T=T_{turn}-T_{max}$ as a function 
of $|f|$. The different points
indicate different values of $|k|$ between $0.5$ 
and $1.5$ Kelvin.
The line is a linear fit with a slope of 0.85 }

As the ESS is the more realistic model the deviation
$T_{max}\not= T_{turn}$  should be visible in
experiments. This will be discussed in the following. 

Note that Fiory \cite{Fiory} has shown that 
even for low concentrations
the 'Schottky anomaly' of the specific heat gets distorted
by the dipole--dipole--interaction between the defects,
and a similiar effect is expected for the dielectric 
susceptibility.
Both effects - the interaction and face tunneling -
 might interfer in a real sample. But by
examing probes with different defect concentrations
it is in principle possible to distinguish between the two
effects; any change depending on a variation of 
concentration results from interactions while effects
that are independent from such a variation arise due to 
face diagonal tunneling.

\subsection{Experimental Data}

Let us now take a look at the experimental data. We
will make use of the specific heat data from Pohl et
al. \cite{Pohl} and the dielectric susceptibility data
from Enss et al. \cite{Christiansus}. Since $f$ will
be the leading correction to the TLS approximation we
will neglect $r$ in the following. We then have two fit
parameters $f$ and $k$ and we can look for the
optimal set in order to reproduce the data.
But indeed there is a third undetermined quantity: the
defect concentration of the probe cannot be
fixed accurately. But the concentration varies only
in a very small range so we focus on the tunneling
parameters. All parameter sets consistent with
experimental data (obtained by fitting the relations
\ref{cvazs} and \ref{schiazs} to the data) 
indeed confirm the relation $T_{max} < T_{turn}$. 

For a further comparison of the two models we will
proceed as follows. 
The TLS is completely specified by the parameter $k_0$
whereas the ESS is specified by the pair $(k,f)$.
Now one can ask: What relation between the ESS 
parameter pair $(k,f)$ and the single TLS parameter $k_0$
must hold in order to produce either the same maximum
of $c_v$ or the same turning point of $\chi$ ?
The $c_V$-maxima of the two models coincide if 
the equation $k+1.7f\approx k_0$ is fulfilled (which 
can be verified by differentiating and solving a
transcendental equation). The turning points of
$\chi$ are the same if $k+2.7f\approx k_0$ holds.
This shows that an TLS-analysis yields indeed
different tunneling parameters for the two
experiments which is counterintuitive;
after all the same degree of freedom produces both
anomalies. Usually such effects were thought to be 
due to experimental uncertainty; but the
above considerations show that the difference is a
systematic consequence of tunneling along the face
diagonal. Based on the difference between $T_{max}$ and
$T_{turn}$ one can estimate an optimal parameter
set. In order to do so one has to solve a set of
transcendental equations, where $T_{max}$ and $T_{turn}$
 are given by the data and $f$ and $k$ are the 
unknown variables.
We have listed the result in table 1.
\begin{center}
\parbox{7cm}{\scriptsize \[  
\begin{array}{|c|cc|}\hline&\mbox{\mas{6}{Li}}& 
\mbox{\mas{7}{Li}} \\ 
\hline \hline
\begin{array}{c} k / k_B \\ 
f/k_B\end{array} & \begin{array}{c} -0.63 K \\ 
-0.08 K \end{array} & \begin{array}{c}  -0.45 K \\ 
-0.06 K  \end{array}  \\ 
\hline 
\begin{array}{c} c_v\; :\; (k+1.7f)/ k_B \\ 
\chi\; :\; (k+2.7f)/k_B\end{array} & \begin{array}{c} -0.77 K \\ 
-0.85 K \end{array} & \begin{array}{c}  -0.55 K \\ 
-0.61 K  \end{array} \\
\hline
\end{array}       
\]\normalsize}
\end{center}
\begin{center}
Table 1: Parameter set consistent with experimental
data. In the lower part of the table effective values
for a TLS are listed.
\end{center}
At first sight the results for $k$ look smaller than 
these reported in the literature.
But keep in mind that most experiments so far were fitted
by a TLS approach.
Therefore we have also listed such 'effective' parameters emerging
from an equivalent fit within the TLS model.
These values are in line with those reported by other authors.
The above parameter set is also in good agreement with
restrictions arising from the WKB-formula (\ref{wkb}).
Then corresponding restrictions concern both the ratio 
\begin{equation} 
\frac{f}{k}=
\exp\left\{
-\frac{\sqrt{2mV_0}}{2\hbar}d(\sqrt{2}-1)\right\} 
\end{equation}
and the ratio of the isotope effect:
\begin{equation}
\frac{\mbox{\mas{6}{k}}}{\mbox{\mas{7}{k}}}=
\exp\left
\{-\frac{d\sqrt{2V_0}}{2\hbar} 
(\sqrt{\mbox{\mas{6}{m}}}-\sqrt{\mbox{\mas{7}{m}}}) 
\right \}.
\end{equation}
The tunneling distance $d\approx
1.4$\AA\ and the mass $m$ of the lithium are known;
the barrier height $V_0$ between all wells should be
of the same order of magnitude. The ratios proposed
above ($f/k\approx 0.13$ and 
\mas{6}{k}/\mas{7}{k}$\approx1.4$) are then consistent with
$V_0\approx 200$ K and $V_0\approx  140$ K
respectively. This seems a rather reasonable value for
the barrier height.
So the above given parameter set is at least consistent
with the given data. We would not like to go further
than this statement,
but only stress, that the ESS is not in 
contradiction with experimental
data. Furthermore it reflects much better the
microscopic picture we have in mind when talking
about a substitutional defect. Still the TLS seems
useful: it has almost the same temperature
dependence of specific heat and static susceptibility 
and is of course much simpler. Nevertheless there
appear some aspects which the TLS cannot reproduce at
all. In the following we will discuss these aspects. In
section 3 such properties arise from the tunneling
parameters $f$ and $r$; we will show that these
features make it possible to measure the tunneling
parameters directly. The properties described in section
4 are based on the particular geometric structure of
the ESS; this structure leads to selection rules which
cannot be understood in the framework of a
one-dimensional model such as the TLS.

\section{Interaction with External Fields}

As mentioned above the defects exhibit both electric
and elastic moments; they  hence couple to the
corresponding external fields. The main effect of
static fields is a modification of the energy levels,
whereas oscillating fields lead to characteristic
transition rules. We will discuss the features in
three steps: first we look at the defect under
pressure and second under the influence of a static
electric field. For both cases we look at the field
dependence and the selection rules for transitions
induced by acoustic and electromagnetic waves. Based
on these results we finally propose an experiment
which can determine the tunneling parameter $f$ (face
diagonal tunneling). Indeed in section 3.1 and 3.2
we will focus on situations where $f$ becomes visible. 

\subsection{The Defect under Pressure}

Let us begin by considering a defect under the
influence of external pressure. As discussed before
(see eq. (\ref{qua})) the defect exhibits an elastic
quadrupole moment
$Q_{ij}\propto\, r_ir_j(1-\delta_{ij})$ which 
interacts with the strain field $\varepsilon_{ij}$. The
latter can be derived by elastomechanical
equations from the pressure exerted on the system
\cite{Landauela}. The interaction energy (cf.
sect. 2) reads
$W_{\varepsilon}=-g\sum_{ij}Q_{ij}\varepsilon_{ij}$. Since
$Q_{ij}$ has (by definition) no diagonal part, the
diagonal part of the strain field is irrelevant for
the interaction. This fact simplifies the
discussion considerably: the diagonal part of the 
strain field depends on the pressure in a rather
non-trivial way.

The energy eigenvalue problem of a defect in an
arbitrary strain  field (i.e. uniaxial pressure in an
arbitrary direction) is not analytically  solvable.
Yet some constellations with high symmetry can be
solved by means of group theory. Such tractable cases
appear for uniaxial pressure in [100]--, [110]-- and 
[111]--direction. Let us write down the
interaction hamiltonian for these cases. In order to do
so one has to derive the strain field which arises from
the exerted pressure \cite{Landauela}. For the
cases we consider these strain fields and the
corresponding interaction energies are listed in 
table 2. 

{\scriptsize
\begin{center}
\parbox{11cm}{\[ 
\begin{array}{|c|c|c|} \hline \mbox{direction of} &
\mbox{distortion field \boldmath$\varepsilon$} & \mbox{interaction}
\\ \mbox{uniaxial pressure} & & \mbox{energy }W_{\epsilon} \\ \hline \hline  
[001] &
\varepsilon_{ij} = \left( \begin{array}{ccc}  \varepsilon_1 & 0
& 0 \\ 0 & \varepsilon_2 & 0 \\ 0 & 0 & \varepsilon_2
\end{array} \right) & 0 \\ \hline  [011] & \varepsilon_{ij} =
\left ( \begin{array}{ccc}  \varepsilon_1 & 0 & 0 \\ 0 & 
\varepsilon_2 & \varepsilon \\ 0 & \varepsilon & \varepsilon_2
\end{array} \right ) & -\frac{4 g \varepsilon}{d^2} (1\otimes
\sigma_z\otimes \sigma_z) \\ \hline  [111] & \varepsilon_{ij} =
\left ( \begin{array}{ccc} \varepsilon_1 & \varepsilon &
\varepsilon \\ \varepsilon &  \varepsilon_1 & \varepsilon \\
\varepsilon & \varepsilon & \varepsilon_1 \end{array} \right ) &
\begin{array}{r}                   -\frac{4 g \varepsilon}{d^2}
\, ( \; 1\otimes \sigma_z\otimes \sigma_z  \\ + \sigma_z \otimes
1\otimes \sigma_z  \\+ \sigma_z \otimes \sigma_z \otimes 1)
\end{array} \\ \hline \end{array} 
\]} 
\end{center} }
\normalsize

Table 2: Distortion field and interaction energy
for different directions

The off-diagonal elements $\varepsilon$ of the strain
fields are proportional  to the pressure;
the proportionality factor depends on the crystal's
elasticity modulus. The case of exerting pressure in
the [100]--direction turns out to be trivial since
the strain field then has no off-diagonal elements and
the interaction energy thus vanishes. The eigenvalue
problem of the other two cases have been discussed in
a two-level-approximation (i.e. $f=r=0$) by Gomez et
al. \cite{Gomez}. We want to avoid this
approximation and focus on the case where the pressure
is exerted in the  [110]--direction (a general
discussion can be found in \cite{Dipl}). In this
situation the Hamiltonian of a defect under uniaxial
pressure reads
\begin{equation} 
H = H_0 -\frac{4 g \varepsilon}{d^2}
(1\otimes\sigma_z\otimes \sigma_z).
\end{equation}
We will not write down any formulas concerning the
eigenvalues and -vectors (which can be found in
reference \cite{Dipl}); instead we will plot all
interesting information and give a discussion. 

Figure 3 shows the dependence of the energy levels on
the pressure. The energies E3, E4, E5 and E6 grow
linearly with pressure; the others vary quadratically
for small pressure and then end up in a linear regime.
All degeneracies are obviously lifted. With growing
pressure the spectrum changes from four almost
equidistant levels to four dubletts; the energy
difference inside such a dublett is asymptotically of
the order of $f$. The upper two dubletts are separated
by a gap from the lower two; this gap varies
(asymptotically) linearly with pressure whereas the
distance of the dublett splitting approaches $\sim2k$. 

{\leavevmode
\centering
\epsfxsize=7cm
\epsffile{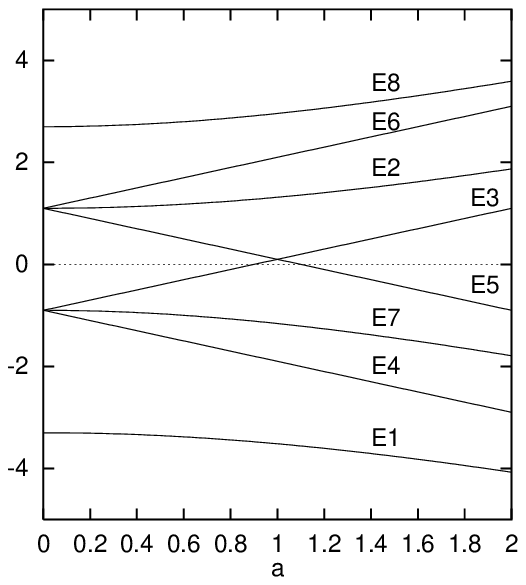}

Figure 3: Energy levels of an ESS (in units of $k$) 
versus $a=4g\varepsilon/d^2$ which is proportional 
to the uniaxial pressure in  [110]--direction}

{\leavevmode
\centering
\epsfxsize=12cm
\epsfysize=4.5cm 
\epsffile{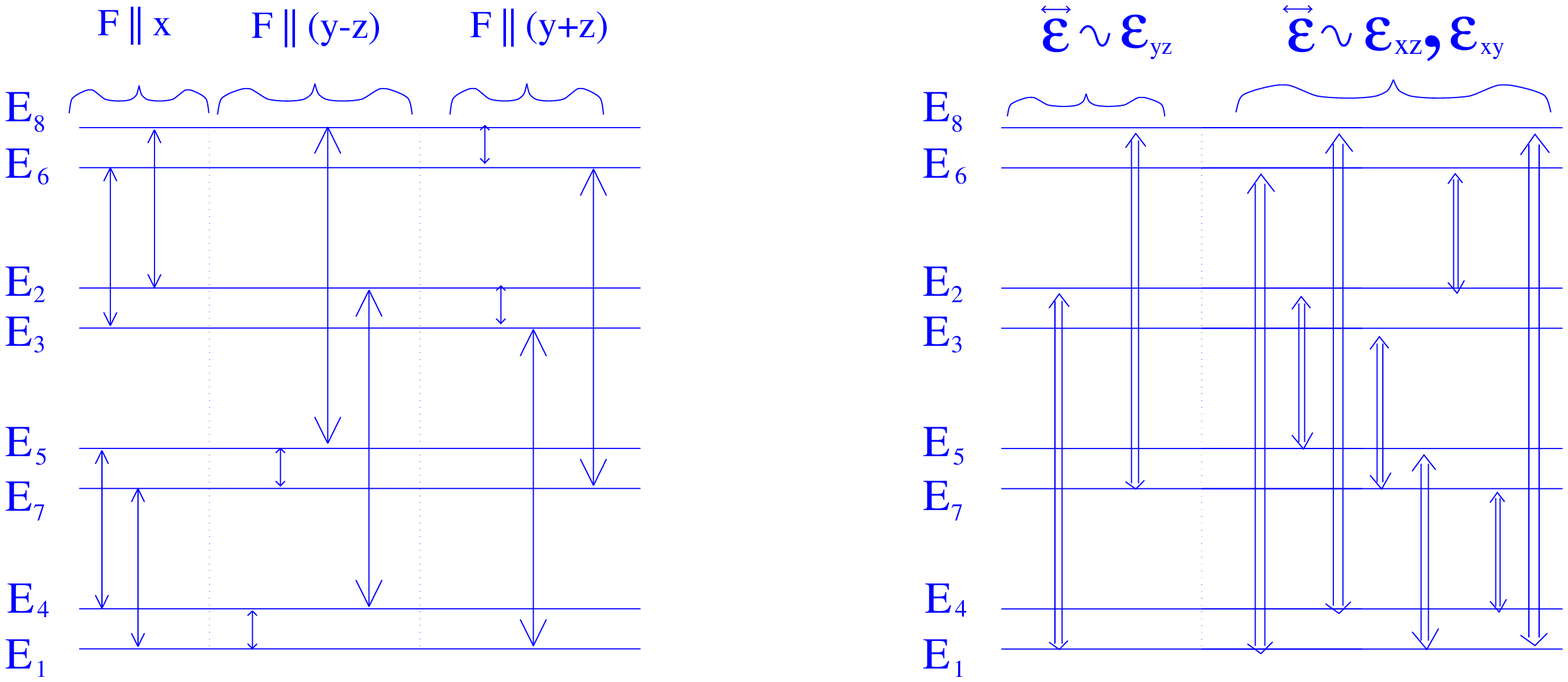} 

Figure 4: Energy levels of an ESS with uniaxial 
pressure in [110]--direction
and  the electric (left picture, single arrows) 
and acoustic (right picture, double arrows) transitions}

Figure 4 shows the possible electromagnetic and
acoustic transitions; the pressure is chosen such
that the coupling energy is roughly the same as the
tunneling energy. In the left part of the scheme
simple arrows indicate a dipole transition
(microwaves) and in the right part double arrows denote
quadrupole transitions (acoustic waves). The symbol
$\vec{F}$ here stands for the electric field of the
electromagnetic wave and {\boldmath$\varepsilon$}
for the (symmetric) distortion field of the 
acoustic wave (with $\epsilon_{ij}=1$ for row index $i$
and column index $j$). The scheme shows that most
transitions can be induced by an appropriate choice
of the direction of the oscillating fields; in
particular the frequency $\sim f$ can be induced by a
microwave field with $\vec{F}$ oscillating in the
[110] direction.

\subsection{The Defect in the Presence of a Static
Electric Field}

Let us now consider the defect in a static electric
field (cf. eq. (\ref{dip})). The coupling energy
between the field and the defect reads
$W_F = -\sum_i F_ip_i$. Again the general situation of a
defect in an electric field of arbitrary direction is
not analytically solvable. Two possible restrictions
allow such a solution. One of them is to neglect $f$
and $r$; in this 'two-level-approximation' the
Hamiltonian
$\hat{H} = \hat{H}_0 - \vec{F} \vec{r}$ factorizes
into three two-dimensional problems.
%
%\bgg\label{hammm} 
%H & =  & \;\:\;\; 1\otimes 1\otimes
%(k\;\sigma_x -  q\frac{d}{2} F_z\;\sigma_z) \\
%       & &     +\;  1\otimes (k\;\sigma_x -
%q\frac{d}{2} F_y\;\sigma_z) \otimes  1 \\
%       & &     +\;  (k\;\sigma_x - q\frac{d}{2}
%F_x\;\sigma_z) \otimes 1\otimes  1 . 
%\egg
%
It is then easy to derive the spectrum and
the transitions. This has been done by Gomez et al.
\cite{Gomez} for special symmetries, but it is in
fact possible for arbitrary directions of the
electric field. The second possibility is to consider
cases where the electric field points along the
[100]--, [110]-- and  [111]--direction. The resulting
problem then still has a high symmetry and group
theory makes it possible to find a solution even for
finite $f$ and $r$. The interesting new feature
emerging is the occurrence of transition frequencies
proportional to $f$ (neglecting $r$) in the [110]--
and  [111]--cases. These transitions are interesting
since they allow a direct measurement of $f$ (which
will be discussed below). We will consider the
[111]--case here; the others can be found in reference
\cite{Dipl}.
\par
The Hamiltonian for a defect with an electric field in
[111]--direction reads
\bgg
 H_{(111)} & = \quad &  k\;( 1\otimes 1\otimes
\sigma_x + 1\otimes \sigma_x \otimes 1 + \sigma_x
\otimes 1\otimes 1 ) \\  &\quad + & f\;( 1\otimes
\sigma_x \otimes \sigma_x + \sigma_x
\otimes 1\otimes \sigma_x + \sigma_x \otimes \sigma_x \otimes 1)
\\  &\quad + & r\;( \sigma_x \otimes \sigma_x \otimes \sigma_x )
\\ & \quad - & q\frac{d}{2} F_{stat}\; ( 1\otimes 1\otimes
\sigma_z + 1\otimes \sigma_z \otimes 1 + \sigma_z \otimes
1\otimes 1 ) . \egg
The presence of the exterior field breaks the cubic
symmetry ($O_H$) of the defect. In the case
considered one is left with the so-called
$C_{3v}$--symmetry group (i.e. there is one threefold
axes and three vertical reflection planes). Group
theory then shows that the spectrum consists of four
singletts and two dubletts since the eight-dimensional
representation decomposes into four one-dimensional
and two two-dimensional irreducible representations.
Correspondingly it is possible to blockdiagonalize the
Hamiltonian. These blocks can then be treated in 
(degenerate) perturbation theory since $f,r\ll k,pF$.
Again we refrain from writing down details and instead
restrict ourselves to a discussion of the results. In
figure 5 the eigenvalues are plotted against the
static field $F_{stat}$. In contrast to the level
shift due to pressure there is no linear field
dependence of any energy level. Neglecting $f$ and $r$
the spectrum conserves the $1:3:3:1$ degeneracy scheme.
The four levels simply spread with increasing field as
$2\sqrt{k^2+(qdF/2)^2}$. Considering finite $f$ and
$r$ changes the situation. The most remarkable new
feature is the splitting of the threefold degenerate
states into a singlett  ($E_1$,$E_2$) and a dublett
($E_5$,$E_6$). This splitting is proportional to the
face diagonal tunneling parameter $f$; it is given by
\begin{equation} E_1 - E_5 = E_2 - E_6 =  f \,
\frac{3 (pF/\sqrt{3})^2}{(p F/\sqrt{3})^2 + k^2}.
\end{equation}

\begin{figure}[h] 
{\leavevmode
\centering 
\epsfxsize=7cm
\epsffile{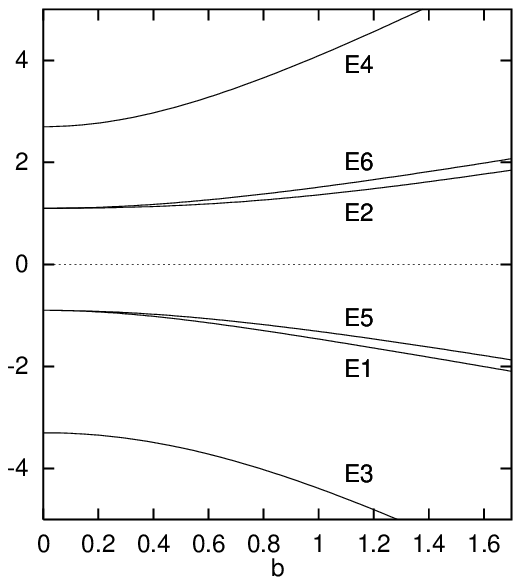} }

Figure 5: Energy levels of an ESS (in units of $k$)
versus $b=pF/\sqrt{3}|k|$ which is essentially 
the static electric  field in [111]--direction
\end{figure}
\begin{figure}[h]  
{\leavevmode
\centering
\epsfxsize=12cm
\epsfysize=4.5cm 
\epsffile{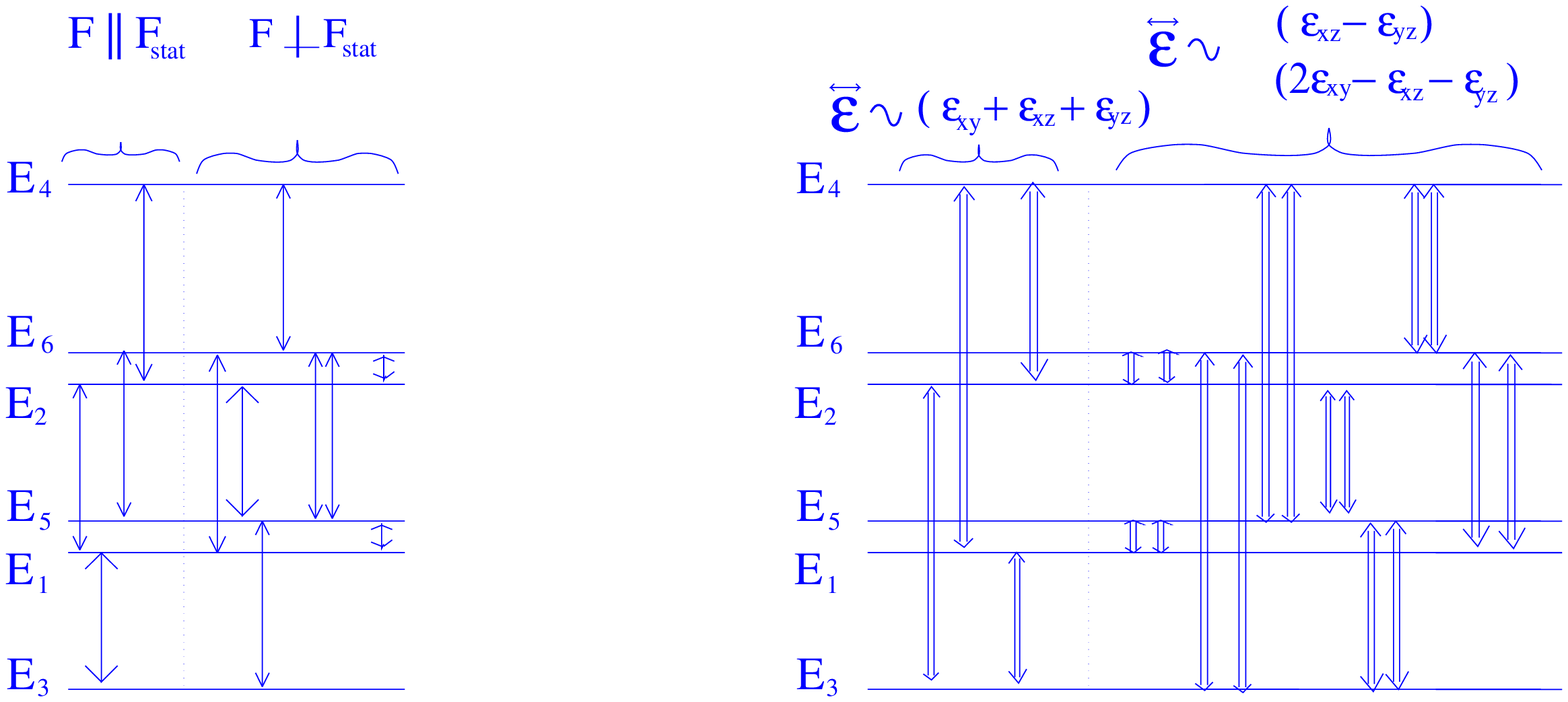} }

Figure 6: Energy levels of an ESS with electric field
in [111]--direction and  the electric (left picture, single arrows)
and acoustic (right picture, double arrows) transitions
\end{figure}

Figure 6 gives the transition scheme of the
situation. In the left part of the scheme we again
show the dipole transitions. Two cases are listed:
one where the electric field of the microwaves
oscillates in line with the static field and one
where it oscillates in a perpendicular direction.

Suppressing amplitudes $\propto\vert f/k\vert^2$ we
find in the first case four transitions:  the selection
rules remain essentially unaltered and lead to
transitions of the order of $\sim 2k$. If the
oscillating field is perpendicular to the static field
many transitions are possible: those 'over the
gap' with a frequency growing with increasing static
field as well as those taking place 'inside' a triplett.
The frequency of the latter is asymptotically
determined by $f$. The probability for the
dipole transition grows with the amplitude of the
static field $\propto F_{stat}^2$. 
\par
The acoustic selection rules are shown in the right
part of figure 6. There is a whole variety of possible
transitions; again we want to focus on the one with
frequency $f$. The distortion tensor
$\varepsilon_{ij}$ for such a transition has to be
proportional to 
$a_1 (2\epsilon_{xy}-\epsilon_{xz}-\epsilon_{yz}) +
a_2  (\epsilon_{xz}-\epsilon_{yz})$ with arbitrary
$a_i$. One possible choice is a sound wave propagating
along the $(y-z)$-direction being polarized
in the $x$-direction. The transition amplitude of
such a constellation is given by 
$\left| \frac{k^2-2(qdF_{stat}/2)^2}
{k^2+2(qdF_{stat}/2)^2}\right|^2$.
The increasing static field reduces the amplitude which
vanishes for $F_{stat}=\sqrt{2} kqd$. For coupling
energies greater than the tunneling energy the
transition probability then again increases. 

\subsection{Measuring Face Diagonal Tunneling}

The results of the preceding sections show that
there are different possibilities to measure the face
diagonal tunneling parameter by choosing an
appropriate geometric constellation for the exterior
static and oscillating fields. Let us focus in the
following on the situation of a static electric field
pointing along the [111]--direction (cf. section 3.2).
The electric field of the microwave is assumed to
oscillate perpendicular to the static field (for example 
in the [0,1,-1]--direction). The selection
rules tell us that a transition with frequency 
$\sim f$ (which should be about 100mK) is possible.
In order to have a transition rate of the order of
unity, one has to apply a static field which leads to
an interaction energy of the order of $k$. This is
fulfilled if $pF\approx k$. Since the dipole
moment of the defect is known to be $p\approx2.6D$
and $k\approx 1$K (cf. \cite{Aloisbuch}), the field strength
should be $F_{stat}\approx 10^6-10^7$V/m. To our
knowledge such an experiment has never been performed.

\section{Strongly Coupled Pairs}

In section 2 we argued that a substitutional defect
exhibits an electric as well as an elastic moment.
These momenta lead to a coupling of the defects
to external fields; moreover they produce an
interaction between neighboring defects. Here again
both are present, the quadrupole coupling of the
elastic moments and the dipole-dipole-coupling of the
electric moments. In potassium chloride doped with
lithium the electric interaction dominates the elastic
coupling by an order of magnitude. Hence only the
dipole--dipole interaction
\begin{eqnarray}
\label{ene} 
&W =\frac{1}{4 \pi \varepsilon \varepsilon_0 }  
\bigg (
\frac{\vec{p_1}
        \vec{p_2}}{R^3} - 3 \frac{( \vec{p_1} \vec{R} ) ( \vec{p_2}
        \vec{R} )}{R^5} \bigg ) = \frac{J}{4} \bigg ( \vec{e}_1 \vec{e}_2 - 3 (
\vec{e}_1 \vec{e}_R ) ( \vec{e}_2 \vec{e}_R ) \bigg ) & 
\end{eqnarray}
with 
\begin{equation} J = \frac{1}{\pi \varepsilon \varepsilon_0 R^3} \frac{p^2}{3}   
\end{equation}
is relevant. Here $\vec{p}_i$ denotes the dipole moment
of defect $i$, $\vec{R}$ is the vector connecting the
defects, the $\vec{e}$ 's are $\vec{e}=(2/d)\vec{r}$,
 $\varepsilon_0$ and $\varepsilon$ are the
dielectric constant of the vacuum and the  potassium
chloride matrix respectively.
\par
The structure of the host crystal confines the defects
to discrete sites on the lattice. Hence the possible
distances between the pairs will be a discrete set.
Although this fact is irrelevant for the bulk of the
pairs which have a distance of, say, ten lattice
constants or more (the crystal then appears as a
quasicontinuum), it is most relevant for the
strongly coupled pairs which we will consider here.
Geometric considerations show that on the fcc
lattice the nearest neighbors NN1 lie along the
[1/2,1/2 ,0]-directions and the next nearest neighbors
NN2 along the [1,0,0]-direction. Comparing the energy
scales for neighboring pairs one finds that the
interaction energy dominates the ESS tunneling
parameters. This can be estimated by evaluating the
dipole--dipole interaction (\ref{ene}) using the lattice
constant $a\approx  6.23$\AA\ and the dipole moment
$p\approx2.6D$. For the nearest neighbors NN1 one is lead 
to a value of $\varepsilon J \approx 710$K (evaluated with
$R=a/\sqrt{2}$). In order
to determine the value of the coupling one has to specify
the dielectric constant of the material. Yet on the
 atomic length scale it is not clear 
which value to take for this constant; at least one can 
confine it to the interval 
$\varepsilon$(vacuum)$=1<\varepsilon <\varepsilon$(KCl)$=4.25$.
This tolerence leads to values $170\, K<J_{NN1}/k_B<710\, K$.
For the next nearest neighbors NN2 the interval is 
$60\, K<J_{NN2}/k_B<250\, K$.
This estimation, as rough as it is, shows that the coupling
between neighboring defects exceeds by at least one order of
magnitude the intrinsic energy scales (which are given by
$k\approx 1$K). The strongly coupled pairs hence 
constitute a composite degree of freedom which 
should be considered as a unit.

Klein \cite{Klein} and one of the authors
\cite{Orestispaper} have discussed defect pairs in the
two-level-approximation; within this approximation
one can predict the existence of a small frequency
$4k^2/J$ for strongly  coupled pairs. For nearest or
next nearest neighbors this energy should be between
5mK and 50mK. Weis et al. \cite{paperrobert}
investigated the system experimentally by measuring
Rabi frequencies in spin echo experiments. He found a
broad distribution of frequencies in the 10mK region;
in addition there was a rather narrow distribution of
Rabi frequencies 
$\Omega_R \propto \frac{4k^2/J}{E}\,\vec{F}\vec{p}$. The analysis
of these experiments was first done by considering
the strongly coupled pair as an effective TLS with a
tunneling rate $4k^2/J$ and an asymmetry $\Delta$.
The total energy of such an effective TLS then reads $E=\sqrt{\Delta^2 +
(4k^2/J)^2}$ (which is the resonance energy of the
external field). Weis now argued that the
measured frequency distribution is due to a
distribution of the asymmetry $\Delta$. Such a distribution
arises from imperfections of the probe, whereas there is no 
plausible reason for a distribution of the dipole moment.
\par
Still the existence of only one frequency was somehow
striking; obviously only one pair constellation
contributed to the signal (otherwise additional Rabi 
frequency peaks would appear). In order to explain
this puzzle W\"urger \cite{Aloisbuch} discussed a
pair of ESS using perturbation theory. He showed that
only the pair constellation NN2 has a frequency in the
range  between 5mK and 50mK. For all other
constellations geometric selection rules prohibit
transitions with  these energies. He also showed that
an asymmetry proportional to the position operator
explains  the distribution of frequencies. Yet he left
open the question which role other forms of
asymmetries (e.g. a 'quadrupole asymmetry' proportional
to $r_ir_j$ which couples to internal strain fields)
will play.

In order to treat this problem one can discuss the NN1
and NN2 case applying group theory and tensor
factorization. In
this context all types of asymmetries are easily
discussed considering them  as small perturbations. We
restrict ourselves to a brief overview of the proposed
methods and present the results for the NN2
constellation (for details cf. \cite{Dipl}). 

First the basis $\{\,|xyz\rangle\,\}$ for one defect
is expanded to $\{\,|x_1x_2y_1y_2z_1z_2\rangle\,\}$
for the pair. In this basis the Hamiltonian reads 
\begin{equation} 
H= H_0 + W 
\end{equation} 
where $H_0$ stands for the tunneling energy of both
defects and $W$ denotes the dipole interaction.
This (64-dimensional) Hamiltonian indeed factorizes
just as in the case of a single defect (we write $k_i$
for the edge tunneling rate of the defect $i$):
\begin{eqnarray}\label{hamnn2} 
H & = & \quad 
\Big ( \,
k_1 (\sigma_x\otimes 1) + 
k_2 (1\otimes\sigma_x) - \frac{J}{2} (\sigma_z\otimes\sigma_z) 
\, \Big )
\otimes {\bf 1}_{4\otimes 4}
\otimes{\bf 1}_{4\otimes 4} 
\nonumber \\      &   & + 
{\bf 1}_{4\otimes 4} \otimes
\Big ( \, k_1 (\sigma_x\otimes 1) + k_2
(1\otimes\sigma_x) + \frac{J}{4} (\sigma_z\otimes\sigma_z) 
\, \Big ) 
\otimes {\bf 1}_{4\otimes 4}
 \\       &   & + 
{\bf 1}_{4\otimes 4}
\otimes{\bf 1}_{4\otimes 4}
\otimes \Big ( \,
k_1 (\sigma_x\otimes 1) + k_2 (1\otimes\sigma_x) + 
\frac{J}{4} (\sigma_z\otimes\sigma_z) 
\,\Big ) .
\nonumber 
\end{eqnarray}
We are left with three four-dimensional problems.
After this factorization it is easy to derive the
complete spectrum together with all selection rules.
This spectrum consists of different multipletts which
are separated by gaps of the order of $\sim \frac{J}{4}$. For
the experiments considered the temperature is much
lower than this energy and hence only the lowest level
group is of interest. In the middle of figure 7 the
spectrum of this lowest multiplett is shown together
with the dipole selection rules. It consists of eight
states with a degeneracy scheme $1:1:2:2:1:1$. The
energy eigenvalues are typically the sum of three
roots of the form 
$\sqrt{(J/4)^2 + (k_1\pm k_2)^2}$ and 
$\sqrt{(J/2)^2 + (k_1\pm k_2)^2}$.

In order to determine the transitions caused by an
external field one has to consider an interaction
with both defects: 
$W_F=-\vec{F}(\vec{p}_1+\vec{p}_2)$. The spatial
dependence of $F$ is neglected since the wavelength of
$F$ ($\approx 1$cm) is much greater than the distance
of the two defects ($\approx 10$\AA). There is only
one transition frequency (just as in the case of a
single defect). But here transitions are only
induced by electric fields $\vec{F}$ oscillating in
line with  the distance vector $\vec{R}$.

The structure of the spectrum together with the
selection rules makes it again possible to talk about
the pair as an effective TLS with tunneling rate 
$4k_1k_2/J$. In particular it is possible to use the
well-known Rabi-formalism for TLS in order to interpret
the echo experiments. In a way this is an a posteriori
justification of the TLS approximation which had been
proposed in the context of defect pairs
\cite{Orestispaper}.

Considering now an asymmetry is somewhat more
complicated as in the two-level-approximation.
There the asymmetry is simply proportional to the
one-dimensional position operator. The ESS is in a
sense three-dimensional and one has to find a way to
assign a different potential value to each well. This
can be done by introducing three dipole terms, three
quadrupole terms and an octupole term. Allowing
different asymmetries at both defects one gets:
\begin{eqnarray}
 V_d & = & \sum_{i=x,y,z} v_i (r_i^1 + r_i^2) +
\delta v_i  (r_i^1 - r_i^2) \\
  V_q & = & \sum_{i,j=x,y,z \atop i\not{=}j} v_{ij} 
(r_i^1r_j^1 + r_i^2r_j^2) + 
\delta v_{ij} (r_i^1r_j^1 - r_i^2r_j^2) \\
  V_o & = &  v_{xyz} (r_x^1r_y^1r_z^1 + r_x^2r_y^2r_z^2) + 
\delta v_{xyz} 
(r_x^1r_y^1r_z^1 - r_x^2r_y^2r_z^2),
\end{eqnarray}
with the mean asymmetry $v=(v_1+v_2)/2$ and the
difference $\delta v=(v_1-v_2)/2$. The existence of
dipole and quadrupole asymmetries is plausible since
it can be deduced from electric and elastic internal
fields which couple weakly to an arbitrarily chosen
defect. In contrast the octupole term has no such
explanation and we hence neglect it in the following.
The asymmetries $V_d$ and $V_q$ will be of the same
order of magnitude as $4k_1k_2/J$. Treating them
properly one is led to the following conclusions:
\par
(i) We find that the part of the dipole mean asymmetry
which is parallel to the distance vector $\vec{R}$
alters the frequency to $\sqrt{(4k_1k_2/J)^2 + v^2}$; 
this result is consistent with that proposed by
W\"urger \cite{Aloisbuch}.
\par
(ii) All other terms (such as the quadrupole terms)
do not change the transition frequency; neither does
the difference in asymmetry.

These results are shown in figure 7. In the middle we
show the lowest multiplett without asymmetry; on the
right hand side we show the 'dipole asymmetry'
changing the energy levels, on the left hand side we
show the 'quadrupole asymmetry' simply shifting the
levels without changing the frequencies of the allowed
transitions. The experiments by Weis are hence only
sensitive to the {\it mean dipole asymmetry}. All
other forms of asymmetry do not play a significant
role here.

{\leavevmode 
\centering
\epsfxsize=7cm
\epsfysize=6cm 
\epsffile{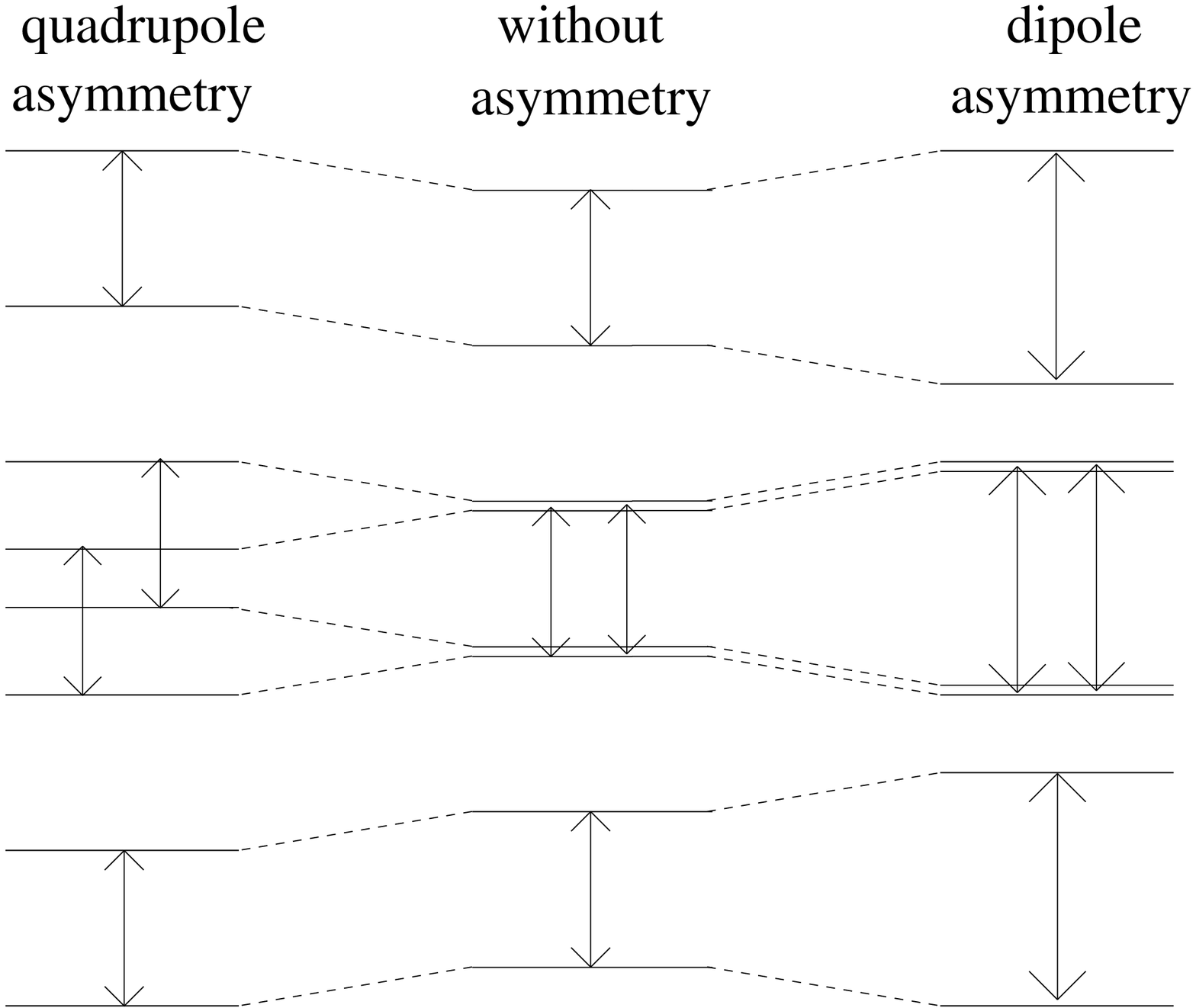}

Figure 7: The lowest multiplett of NN2}

\section{Conclusion}

In this article we had the aim to compare the physics
of a two-level-system (TLS) to that of an
eight-state-system (ESS). Both systems are used
as possible models for cubic substitutional defects
such as the lithium atom in a KCl host lattice. The
ESS reflects much better our microscopic picture of
the defect and it is hence interesting to know what
features are left out of consideration when treating
the defect as a TLS.

Using a tensorial notation we reminded the reader that
the ESS reduces to three TLS if tunneling along the
face and space diagonal of the cube is neglected.
Then we presented three contexts where differences
between the two models arise. First we looked at the
temperature dependence of the specific heat and
the static dielectric susceptibility. A simple way to
summarize the differences of the models is to state
that for the TLS the relation $k^2\,\partial_T\schi\;
=\; c_v$ holds, whereas this is not true for the ESS.
In particular the maximum temperature of specific heat
and the turning point of the susceptibility coincide
for the TLS, whereas they are different for the ESS.
We think that the experimental data indeed shows this
tendency. The ratio  $f/k$ (face to edge tunneling)
and the isotope effect \mas{6}{k}$/$\mas{7}{k}
estimated from the data leads to reasonable orders of
magnitude for the barrier height (100-200K).

Second we investigated the differences of the two
models in the presence of external forces (uniaxial
pressure and static electric field). There appear
different aspects where tunneling along the face and
space diagonal of the cube is visible. These
features make it possible to set up a resonance
experiment where the face diagonal tunneling can be
measured  directly: applying the electric field along
the [111]--direction, there are (both acoustic and
electromagnetic) transitions with frequencies $\omega
\sim f (pF)^2/((pF)^2+k^2)$ (where $p$ denotes the
dipole moment and $F$ the static electric field). Of
course such features cannot be described within the
framework of a TLS.

As a third context we discussed strongly coupled
defect pairs. Here again some properties arise that
cannot be explained by a two-level-approximation. For
strongly coupled pairs {\it geometric} selection
rules inhibit transitions for some constellations
(such as the nearest neighbors). In addition
different asymmetry terms are possible for an ESS,
whereas in the TLS only one such term appears. It is by
the interplay of selection rules and the structure of
the spectrum, that these other asymmetries are not
observable in the echo experiments performed. This is
why the situation can be described in terms
of an effective TLS. 

We thus conclude that there is a number of situations
where the ESS shows observable features which cannot
be described by a TLS. Nevertheless the latter is a
good approximation in all contexts where the geometry
and the parameters for tunneling along the face and
space diagonal are of no relevance. In these
situations the TLS being the simplest model for a
tunneling degree of freedom indeed sketches all
essential features of the defect.

\vspace{0.8cm}
{\bf ACKNOWLEDGEMENTS}
We wish to acknowledge helpfull discussions with C. Enss,
H. Horner, B. Thimmel and A. W\"urger.

\end{document}